# Experimental Comparison of SNR and RSSI for LoRa-ESL Based on Machine Clustering and Arithmetic Distribution

Malak Abid Ali Khan[1], Hongbin Ma[*2], Syed Muhammad Aamir[3], Cekderi Anil Baris[4]

School of Automation, Beijing Institute of Technology, Beijing 100081, China
mathmhb@bit.edu.cn

**Abstract.** LoRa lacks the sensing capabilities of channel status. Received signal strength indicator (RSSI) decreases due to collision, interference, and near-far effect while for signal-to-noise ratio (SNR), the packets are rejected by decreasing the transmission power (TP) at a higher spreading factor (SF). To overcome these challenges in the case of electric shelf label (ESL) to minimize the dependency on retransmission and acknowledgment, the end devices (EDs) are allocated around gateways (GWs) based on machine clustering with dynamic SF for SNR while dynamic TP for RSSI. The experimental results determined that the RSSI approach is more dominant than SNR because of determining the exact locality of the ED that diminished the capture effect. Arithmetic distribution of EDs for various GWs in different clusters helps to minify the near-far effect. The resultant received power (RP) at each cluster is higher for most of the connected EDs than the threshold RP.

## 1. LoRa & LoRaWAN

LoRa is one of the most widely used low-power wide-area networks (LPWANs) in the current era for the building internet of things (BIoTs). It uses Chirp Spread Spectrum (CSS) modulation mechanism to modulate signals. A chirp sweeps through and wraps around a predefined bandwidth, referred to as up-chirps and down-chirps by using CSS which refers to a signal with constantly increasing or decreasing frequency. LoRa configuration may be changed by manipulating some key parameters to achieve change-offs amongst verbal exchange distances, information charges, and electricity consumption. In addition to controlling facts charge, the range of spreading aspect choices additionally increases the co-existence of LoRa devices, i.e., one kind of SFs permits the demodulator to broaden a component of resilience in opposition to simultaneous transmissions of various SFs within the same channel and demodulate it all. This type of characteristic significantly enhances the multiple gets right of entry to the efficiency of LoRa [1, 2]. LoRaWAN is designated for powerless and low data communication purposes that can be used directly to communicate with the gateway because it is a lightweight protocol. It specification varies barely from region to place-based totally on the special regional spectrum allocations and regulatory necessities as shown in table 1.

**Table 1** Characteristics of LoRa

| Region | North America | Europe |
|---|---|---|
| Frequency Band in MHz | 902~928 | 867~869 |
| Channels | 64+8+8 | 10 |
| Channel BW Up in kHz | 125/500 | 125/250 |
| Channel BW Dn in kHz | 500 | 125 |
| TX Power Up in dBm | 20~30 | +14 |
| TX Power Dn in dBm | +27 | +14 |
| SF Up | 7~10 | 7~12 |
| Data Rate | 980bps~21.9kbps | 250bps~50kbps |
| Link Budget Up | 154dB | 155dB |
| Link Budget Dn | 157dB | 155dB |

It is extraordinarily crucial for any to have protection. LoRaWAN makes use of layers of security: one for the network and one for the application. Advanced encryption standard (AES) is used with the key trade utilizing an IEEE extended unique identifier (EUI64). The Data Link Layer (DLL) is defined by the LoRa Alliance.

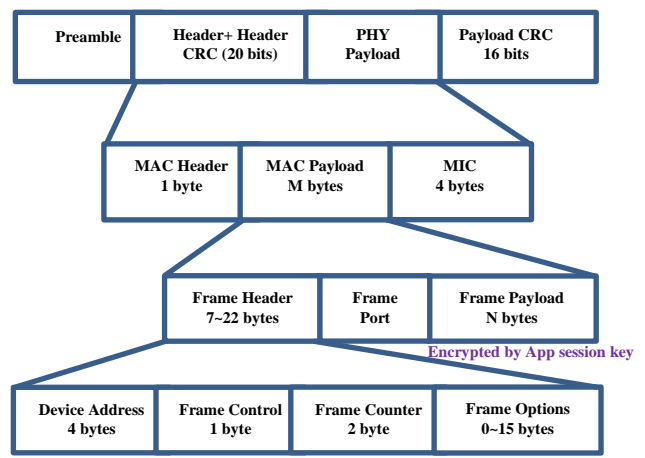

**Fig. 1** Structure of LoRaWAN

To optimize energy consumption, LoRaWAN uses a pure ALOHA medium access control mechanism, encompassing three classes of EDs: Class A, Class B, and Class C. LoRa classes are based on network downlink communication latency and battery lifetime relationships. Downlink communication latency is an important factor in control or actuator-type applications. LoRaWAN protocol defines three classes of EDs based on bidirectional communication, downlink latencies, and power requirements [3, 4].

## 2. ELECTRONIC SHELF LABEL

The ESL consists of an application, communication, and display module. The network server (NS) in the ESL is the hub of star topology and has a great responsibility to optimize the data transmission rate and the energy of the EDs, intending to optimize network scalability and energy consumption with the help of adaptive data rate (ADR). TP and SF are controlled by Network Server (NS). The performance of dynamic ADR is considered by three main strategies, namely dynamic SF, different DR, and allocating least EDs to the clusters based on Arithmetic series. For constant TP, each EDs estimate the best SF configuration depending on the distance from the GW. Its market is constantly evolving because of technological adoptions. The growing variety of companies in the industry has resulted in an expanded number of merchandise, thus creating confusion for most of the clients, even though the maximum of the products is primarily based on the E Ink platform. The manufacturers are investing increasingly more in differentiation based on layout and software. By Technology as shown in Fig. 2 [2, 5], the RF (Radio frequency) ESL label market will grow at a CAGR of around 12% because of the benefit of the use of these merchandise with Wi-Fi, Zigbee, BLE, and Z-wave. Several producers use these technologies because of couple layers of safety in imparting secured low-power wireless merchandise. This allows the shops to combine these devices with the prevailing handheld devices, decreasing the overall installation value of these labels.

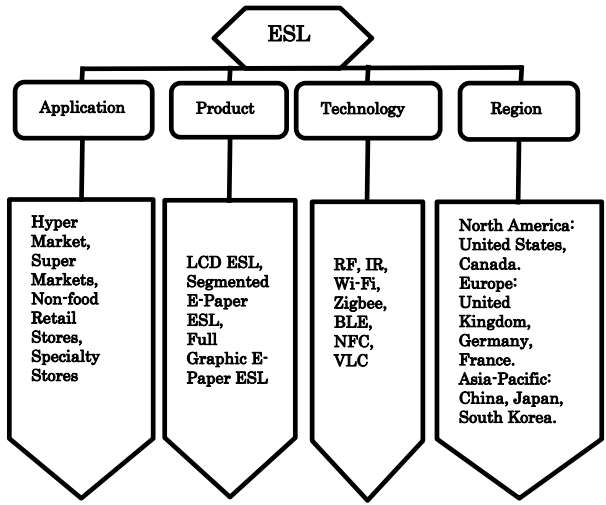

**Fig. 2** Types of ESL

## 3. RELATED WORK

Martin Bor. et.al. [6] have discovered an optimal transmission parameter setting that balances network performance and energy consumption. It was discovered that with 285 probes, a setting that uses just 44% more energy than the ideal setting may be found. The stated method can be added to LoRaWAN, which already has one but doesn't specify how to apply it. Valentina Di Vincenzo presented three strategies to reduce these effects: a multi-gateway implementation, concurrent sending of downlink frames, and a balanced gateway selection mechanism. In the presence of a high downlink load equivalent to 100% confirmed messages, the proposed solutions deliver a 66% improvement in a quad-gateway scenario, compared to an initial 86 percent frame loss with a single-gateway. Furthermore, the balanced gateway selection process, frame loss never surpassed 5% for low and medium downlink traffic loads, allowing the deployment of applications that require such guaranteed data transfer. Carlsson, A. et.al. [7] suggests that large-scale deployments of a gateway on a radio tower to avoid obstructions would have no trouble reaching distances of +15 kilometers. Barriers in the shape of houses, topography, and/or vegetation quickly add up in smaller size implementations, limiting the implementation range to a house or neighborhood in most circumstances. Elevating the entryway to an ideal height may be impossible in some instances. The approach would be to employ many gateways in the same implementation, similar to how a device can connect to any radio tower in a traditional telecommunication system. Another consideration is whether the device is mobile, as misaligned antennas have a significant impact on performance, with packet loss ratio (PRR) differences of up to 50%. In terms of packet loss, it appears that reaching the -120dBm threshold corrupts the packet, rendering it unintelligible or altogether lost. SNR, on the other hand, appears to have less of an impact, as a packet appears to be just as legible at -14dB as it is at a positive value. As the RSSI might fluctuate and there are usually no retransmissions in LoRa, if the device is at the end of the communication range, the packet can be lost forever for most implementations. Andri Rahmadhani. et.al. [8] showed the various aspect of collision in LoRaWAN. The four cases of collision in a single gateway; when a stronger frame arrives later than the receiver locking time. When a stronger frame overlaps with the header of a weaker frame. When the receiver is finishing the header of a weaker frame and a strong frame arrives. The receiver is locked on the weaker frame head while data is inside LoRaWAN. When the receiver is finishing the LoRaWAN header of the weak frame and the strong frame arrives. Dmitry Bankov. et.al. [9] worked on the channel access of LoRa, and evaluate the performance of the LoRaWAN protocol by connecting N nodes to a gateway. For this purpose European configuration was used. The path losses were calculated by the Okamura-Hata model. The collision was considered to occur when at least two transmissions at the same data rate overlap in time. All the physical layer payload was 64 byte and was fitted at the lowest data rate. This work was based on analyzing the collision during multiple nodes and a single gateway of LoRa.

## 4. PARAMETERS SELECTION

### 4.1. Bandwidth

BW represents the width of the frequencies in the transmission band ranging from higher to lower frequencies. Higher frequencies mean higher BW values that provide a higher data transfer rate and greater sensitivity to noise. The sets of BW in LoRa are 125 kHz, 250 kHz, and 500 kHz which determines the width of the transmitted signal. BW determines the chirp duration where each chirp consists of $2^{SF}$ the number of chips. Chip duration is also changing according to the varying BW which also affects the SNR or maybe that particular chirp [10, 11].

### 4.2. Spreading Factor

SF is the ratio between the symbol rate and the chip rate, which can be in the range of seven to twelve. Its level is determined due to the number of bits filled into a chirp. For example, SF7 means that every chirp represents seven bits. As the LoRa chirp needs to be $2^{SF}$ of several starting/ending frequency levels. Receivers would have greater possibilities to sample the signal power, resulting in higher SNR by doubling the chirp duration. A higher value of SF will increase the transmission range of SNR, and also the airtime of the packet. As a result, the data rate will decrease.

### 4.3. Transmission Power

Transmission energy directly affects the quantity of energy used to transmit a chirp. By increasing it, the signal will have a higher probability of surviving attenuation from the surroundings which efficaciously increases the signal power acquired by receivers. Semtech gateway SX1301 has -142dBm sensitivity which can run at an extremely terrible SNR of 9dB. It is capable to operate the transmitted signal that is below the noise ground.

### 4.4. Coding Rate

LoRa modulation provides Forward Error Correction (FEC) to a packet before transmission, imparting safety towards transmission interference by encoding 4-bit data with 5-bit to 8-bit redundancies, allowing the receiver to detect, and accurate errors within the message. LoRa makes use of the Hamming Code because of the forward error correction code utilized in CR. The coding rate (CR) expression is $CR=4/4+n$, and $n$ is from 1 to 4. Better CR values provide greater interference protection.

## 5. NETWORK EVALUATION

### 5.1. Path Loss Model

It measures the path loss of indoor radio channels on Large-scale fading characteristics purposes that can be used for link budget calculation. One-slope model is a type of path loss model that has been determined to perform well in indoor environments [12, 13]. According to the empirical model:

$$L_{PL}(d) = L_{PL}(d_0) + 10n\log(d/d_0) + X_\sigma \qquad (1)$$

Where $L_{PL}$ is the path loss of the reference distance, $n$ is the path loss exponent, $d$ is the separation distance between GW and ED, and $X_\sigma$ is a zero-mean Gaussian distributed variable with standard deviation σ in dB. Table 2 shows the attenuation factors of different obstacles for a 2m distance between GW and ED.

**Table 2** Attenuation Factor

| Obstacles | AF (dB) |
|---|---|
| Concrete wall | 2.2 |
| Glass (2cm) | 2.04 |
| Wooden door | 2.11 |
| Soft partition board | 2.5 |

### 5.2. Near-Far and Capture Effect

EDs and GWs, if a GW is closer to the ED while the others are far away, and all the GWs transmit the signal simultaneously at equal powers, then due to the inverse square law, the ED will receive more power from the nearer GW and this is the near-far effect which helps to detect or filter out a weaker signal among stronger signals. This phenomenon is also called the capture effect where the strongest packet can be received successfully at the receiver side while others will be considered as noise [14].

### 5.3. Machine Clustering

In this work, each iteration defines the set of EDs at the outer cluster. The algorithm seeks the set of $K$ centroids C that minimizes the average of the distances between any ED and its closest GW. Where $ED_n$ is the set of devices at the $K^{th}$ iteration, $ED_t$ is a device in $ED_n$, and $GW_n$ is the closest centroid of $ED_n$ which computes the Euclidean distance between EDs. The EDs in the $C$ boundary form the set of clusters that are separated and act as vectors storing the coordinates of the inner EDs for each Cartesian dimension. The maximum absolute value of each dimension is calculated to set the radius for defining the limit of the cluster [2, 11].

$$C = \arg\min_{GW_n \in C} \frac{1}{|ED_n|} \sum_{ED_t \in ED_n} dist(GW_n, ED_t)^2 \qquad (2)$$

### 5.4. Arithmetic Progression

An arithmetic progression is a sequence of numbers such that the difference between the consecutive terms is constant. The initial term for this progression is $ED_{C1}$ and the common difference of connected EDs is $d$, then $n^{th}$ term of the distributed $ED_{Cn}$ is given for a single GW:

$$ED_{Cn} = ED_{C1} + (n-1)d \qquad (3)$$

For Multi GWs the Arithmetic Distribution is;

$$ED_{Cn} = \{ED_{C1} + (n-1)d\}/GW_n \qquad (4)$$

# 6. ALGORITHMS

## Algorithm 1. SNR

*Input: EDs, GWs, Range.*
*Initialize: t, BW, SF*
*Ensure: ED class*
*If Class A*
*Receive a configuration frame*
*else Class B*
*Receive a beacon from the gateway (ping)*
*for t=7,8,… and TP=14 do*
*Calculate SNR for each ED.*
*Calculate the Margin for each cluster via Eq.5*
*Estimate connected ENs using each GW via Table 3*
*if nStep > 0 and Margin> Threshold do*
    *Send new data packet to ED*
*else if nStep > 0 and Margin < Threshold do*
    *Lower the SF accordingly*
    *drIdx ← drIdx +1*
    *nStep ← nStep -1*
*else*
 *alert unsuccessful transmission*
*end if*
*end for*

## Algorithm 2. RSSI

*Input: EDs, GWs, Range.*
*Initialize: t, BW, TP*
*Ensure: ED class*
*If Class A*
*Receive a configuration frame*
*else Class B*
*Receive a beacon from the gateway (ping)*
*for TP= 14,17,…. do*
*Calculate RSSI for each cluster via Eq. 6*
*Estimate connected ENs using each GW via Table 3*
*if ( Received RSSI > Receiver RSSI) then*
    *GW transmits a new packet to EN*
*else if (Received RSSI < Receiver RSSI) then*
    *TP ← TP+3*
*else*
 *alert unsuccessful transmission*
*end if*
*end for*

# 7. PERFORMANCE & EVALUATION RESULTS

## 7.1. SNR Evaluation Based on Constant Power Transmission Power and Arithmetic Distribution

RSSI and SNR are mathematically related. The reason we choose SNR for this experiment, the same range is used by the radio chipset manufacturer to measure both the signal and the noise. If signal and noise are measured using the same chipset, then SNR is a reliable indicator. More EDs assign to higher SF have more chances of collision and co-SF interference. To overcome this problem, the Fibonacci (increases the chances of far-near effect and collision) and Arithmetic series were used for allocating EDs in different clusters as shown in table 3.

**Table 3** Number of EDs connected per GW

| Range | GW | 2 GWs | 4 GWs | 10 GWs | 20 GWs |
|---|---|---|---|---|---|
| 2.1 | 200 | 100 | 50 | 20 | 10 |
| 1.6 | 600 | 300 | 150 | 60 | 30 |
| 1.5 | 500 | 250 | 125 | 50 | 25 |
| 1.1 | 400 | 200 | 100 | 40 | 20 |
| 0.9 | 300 | 150 | 75 | 30 | 15 |
| 0.7 | 200 | 100 | 50 | 20 | 10 |

To avoid this issue, the Arithmetic series was used that allocates the least number of EDs to the near and far clusters. This approach decreases the saturation in the nearest cluster and collision in the far-most cluster. The following parameters are used for SX1276 in this experiment.

**Table 4** Parameter selection for SNR

| Parameter | Value |
|---|---|
| CF | 868 MHz |
| BW | 125 kHz |
| SF | 7–12 |
| TP | 14dBm |
| Payload | 16 bytes |

Positive values of Margin for demodulation represent strong signals while negative values represent weak signals. LoRa has SNR below the noise floor, and the measured SNR below the required SNR is corrupt [2].

$$Margin = SNR_{Measured} - SNR_{Gateway} \quad (5)$$

A different number of gateways is used to examine the performance of the network for constant TP to avoid packet rejection. The lower SF performs better than the higher SF for indoor activities. GWs and EDs need time to move from the lower to the higher value of SF for maintaining connectivity which results in packets lost. In the case of a single gateway, most of the packets are corrupted which leads to unsuccessful reception at the ED while for two gateways more than half of the packets are delivered successfully. The four gateways perform better than two gateways but still need optimization while the ten gateways perform even better than others. But the network suffers from the far-near effect that leads to unsuccessful reception at far most EDs. Fig. 3 shows the performance of the network for twenty gateways where strong SNR single is received at the reception side for SF12 while weaker at SF11 and SF8. The overall performance of the ten gateways is better and more economical.

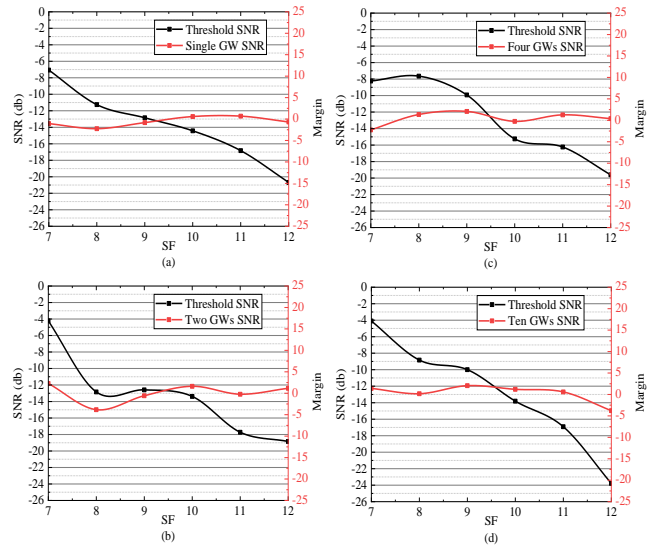

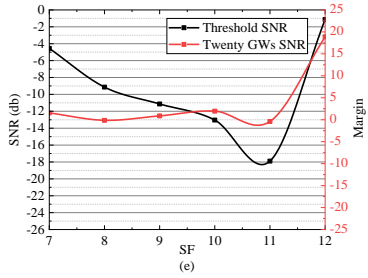

**Fig. 3** SNR performance of each GW in different clusters

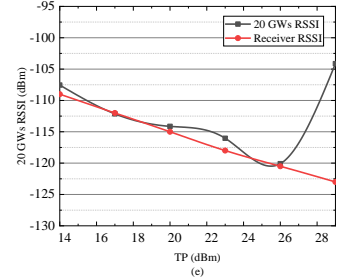

**Fig. 4** RSSI performance of each GW in different clusters

### 7.2. RSSI Evaluation Based on Dynamic Transmission Power and Arithmetic Distribution

RSSI is the strength that ED is hearing a GW/GWs. When the transmissions of two LoRa GWs overlap at the EDs, several conditions determine whether the ED can decode the frames or not. These conditions depend on the channel, SF, TP, and timing [11]. Obstacles and distance between the GW and ED decrease the RSSI. As LoRa is a form of frequency modulation, it presents the capture effect. To overcome this problem, each GW in a specific combination is allotted a constant number of EDs. TP helps to maintain the RSSI accordingly. Low TP is used for the EDs in the nearest cluster while high for the far most EDs in the last cluster. Clustering helps to determine the circumference of each GW which avoids the capture effect. It is rare to suppress any signal of the EDs for the closer clusters of the GW. To build ESL for 2200 EDs, the following parameters are used.

**Table 5** Parameter selection for RSSI

| Parameter | Value |
|---|---|
| TP | 14-29 dBm |
| $L_{PL}$ | 127.84 dB |
| $G_{TX}$ | 2.15 dBi |
| Path Loss Exponent | 4.31 |

$$RSSI_{(dB)} = TP_{(dBm)} + G_{TX\,(dBi)} - L_{PL\,(dB)} \qquad (6)$$

Where TP denotes transmitter power, $G_{TX}$ represents antenna gain of the transmitter, and $L_{PL}$ denotes path loss of the channel, which includes multipath and shadow fading.

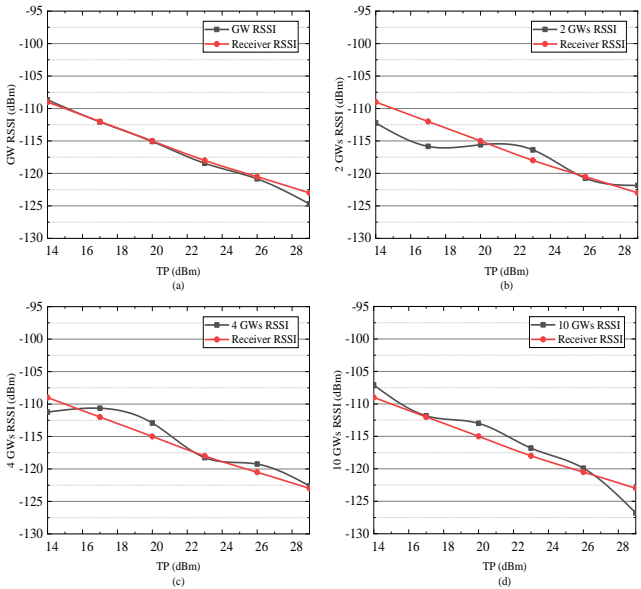

The strong signal RSSI is -104.13dBm while the weak signal RSSI is -126.78dBm in this work. The probability of successful transmission depends on the values of dynamic TP and the allocated EDs using 125 kHz for each GW. An increasing number of ENs drastically diminishes the performance of the network at higher SF values in the case of the signal gateway. As the RSSI for a GW is useful in determining the real position of the EDs in the range of 0.7 km to 2.1 km will change the TP according to the number of the cluster. Receiving signals carrying the same data packet from different GWs causes congestion at the ED. To cover this problem in a multi GWs scenario, Each GW has a defined circumference with a specific number of EDs in each cluster. Each combination of GWs defines the performance of the network as shown in Fig. 4.

### 7.3. Received Power at EDs

The clustering helps to minimize the overlapping of packets from different GWs at the EDs due to dynamic power distribution to avoid the far-near effect. RP then becomes:

$$RP_{(dBm)} = TP_{(dBm)} + G_{TX\,(dBi)} - L_{PL\,(dB)} + G_{RX\,(dBi)} \qquad (7)$$

$$RP_{(dBW)} = RSSI_{(dB)} + SNR_{(dB)} - (1 + 10^{0.1SNR}{}_{(dB)}) - 30 \qquad (8)$$

As from Eq. 6 and Eq. 7,

$$RP_{(dBW)} = RSSI_{(dB)} + G_{RX\,(dBi)} - 30 \qquad (9)$$

Fig. 5 shows that each GW in a different combination has a different RP at the reception side (EDs). The results based on RSSI suggest that RP for a single GW illustrates almost half of the packets are lost due to the maximum number of EDs per GW while for each multi GWs the results are dramatically changed. The variation in the Packet Loss Ratio (PLR) suggests that the multi GWs perform better up to a certain limit and then degrade due to congestion in the near clusters. For two GWs more than 60% of the packets are received with RP more than threshold while for four GWs only 20% of packets are under threshold RP. In the case of ten GWs, each GW follows machine clustering with the Arithmetic distribution of the EDs, all packets received at the EDs with a power higher than the threshold area. The smooth flow of data packets at the near and far clusters determines the equal distribution of power and EDs. The twenty GWs perform better at the far clusters than the nearest due to less number of EDs as compared to other combinations of GWs. The nearest clusters suffer from congestion and capture effects due to the maximum number of GWs in a specific area.

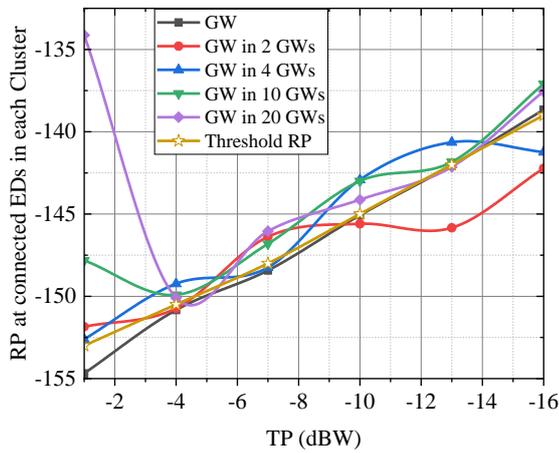

**Fig. 5** RP at each ED in different clusters

## 8. CONCLUSION

The packets get lost due to poor signal quality and are maintaining a mean SNR value but lower RSSI than the sensitivity of the receiver. The attenuation of the signal happens due to passing through or around several materials where most materials are harder than wood inside the shopping mall or warehouse, which means that for BIoTs, dependency on the placement of both GW and ED can have a large effect on the quality of the signal. Regarding packet loss, it looks that exceeding the threshold of -123dBm causes the packet to become corrupted or rejected. SNR seems to have less effect as the packet is readable at ∼-20dB in the case of 20 GWs. This means that for most implementations, it is very important to have packet loss as the RSSI can fluctuate with no retransmissions in LoRa. If the ED is at the edge of the communication range, the packet can be lost forever. This can be very painful for ESL as the duty cycles are very limited and the next transmission can be hours away. We noticed that due to the Arithmetic distribution and dynamic TP of the GWs, each GW offers immunity against multi-path and signal fading, especially at high SFs, when testing network performance for LoRa's RSSI values. The RSS is higher at closer distances to the GW due to the low SF scheme. Furthermore, increasing the SF reduces packet loss at the expense of a lower effective bit rate, which is incompatible with high-throughput BIoT applications. EDs should communicate at high SFs since interferences are substantially higher at the farthest locations from the GWs. Thus, allocating the minimum number of EDs to the last cluster is the best solution to this problem. The RP depends on the TP and all gains and losses along the communication path. On the GW, the range can only be changed by changing the TP. Other parameters like SF and CR do not influence TP, or any other gains and losses. On the ED, the range is limited by the TP threshold, which is influenced by the LoRa parameters SF and BW.

**Acknowledgments**

All the authors sincerely thank the editor, associate editors, and anonymous reviewers for their feedback that enhanced the quality of this work.

**Funding**

This work was partially funded by the National Key Research and Development Plan of China (2018AAA0101000) and the National Natural Science Foundation of China (62076028).